\documentclass[conference, letterpaper, 10pt]{IEEEtran}
\usepackage{cite}
\usepackage[pdftex]{graphicx}
\usepackage[cmex10]{amsmath}
\usepackage[hyphens]{url}

\renewcommand{\baselinestretch}{0.933}

\usepackage[table,x11names]{xcolor}
\usepackage{subcaption}
\usepackage{hyperref}

\usepackage[listings,skins,breakable]{tcolorbox}

\usepackage{newfloat}
\DeclareFloatingEnvironment[fileext=frm,placement={tph},name=Listing]{code}
\DeclareMathOperator{\CFG}{CFG}
\newcommand{\DTW}{\ensuremath{\mathit{dt\_dyn}\xspace}}
\tcbset{enhanced jigsaw,colback=gray!1!white,colframe=black,toprule=0pt,titlerule=0mm,arc=0.2mm,breakable}

\let\labelindent\relax % https://tex.stackexchange.com/a/188255
\usepackage[inline]{enumitem} 


\usepackage{nameref}
\usepackage{varwidth} %for the varwidth minipage environment

\definecolor{commentgreen}{RGB}{176, 176, 176}
\definecolor{rowcolor}{cmyk}{0,0.87,0.68,0.32}
\definecolor{rowcolor2}{cmyk}{ 20, 0, 37, 34}

\definecolor{eminence}{RGB}{108,48,130}
\definecolor{weborange}{RGB}{255,165,0}
\definecolor{frenchplum}{RGB}{129,20,82}
\definecolor{darkgreen}{RGB}{10, 92, 10}

\definecolor{celadon}{rgb}{0.67, 0.88, 0.69}
\newcolumntype{a}{>{\columncolor{rowcolor!40}}r}
\newcolumntype{t}{>{\columncolor{celadon!50}}r}
\newcolumntype{g}{>{\columncolor{gray!30}}r}

\lstdefinestyle{WATStyle}{
  numbers=left,
  stepnumber=1,
  numbersep=10pt,
  tabsize=4,
  showspaces=false,
  showstringspaces=true,
}

\lstdefinestyle{LLVMStyle}{
  numbers=none,
  stepnumber=0,
  numbersep=10pt,
  tabsize=4,
  showspaces=false,
  showstringspaces=true,
}

\lstdefinestyle{CStyle}{
  numbers=none,
  stepnumber=1,
  numbersep=10pt,
  tabsize=4,
  showspaces=false,
  showstringspaces=true,
}
\lstdefinelanguage{WAT}{
    otherkeywords={},
    morekeywords=[1]{i32,f32,i64,f64},
    morekeywords=[2]{0},
    morekeywords=[3]{add,const,mul,shl,get,rem_s,rem_u,ne,tee,sub,set,store},
    morekeywords=[4]{},
    morekeywords=[5]{global, get_global, mut, set_global, export, import,loop, memory, data, get_local,if, block,module, set_local,call,br_if,end, all,call_indirect,local,global,module, func, param, result, type},
    morekeywords=[6]{=,;},
    morekeywords=[7]{(,),[,],.},
    sensitive=false,
    morecomment=[l]{;},
    morecomment=[s]{;}{;},
    morestring=[b]",
    keywordstyle=[1]\color{eminence}\bfseries,
    keywordstyle=[3]\color{frenchplum},
    keywordstyle=[5]\color{darkgreen}\bfseries,
    commentstyle=\color{commentgreen}
}
\makeatletter
\lstdefinelanguage{llvm}{
    morecomment = [l]{;},
    morestring=[b]", 
    sensitive = true,
    morekeywords=[2]{i32,f32,i64,f64},
    morekeywords=[3]{
        define, declare, global, constant,
        internal, external, private,
        linkonce, linkonce_odr, weak, weak_odr, appending,
        common, extern_weak,
        thread_local, dllimport, dllexport,
        hidden, protected, default,
        except, deplibs,
        volatile, fastcc, coldcc, cc, ccc,
        x86_stdcallcc, x86_fastcallcc,
        ptx_kernel, ptx_device,
        signext, zeroext, inreg, sret, nounwind, noreturn,
        nocapture, byval, nest, readnone, readonly, noalias, uwtable,
        inlinehint, noinline, alwaysinline, optsize, ssp, sspreq,
        noredzone, noimplicitfloat, naked, alignstack,
        module, asm, align, tail, to,
        addrspace, section, alias, sideeffect, c, gc,
        target, datalayout, triple,
        blockaddress
    },
    morekeywords=[4]{
        fadd, sub, fsub, mul, fmul,
        sdiv, udiv, fdiv, srem, urem, frem,
        and, or, xor,
        icmp, fcmp,
        eq, ne, ugt, uge, ult, ule, sgt, sge, slt, sle,
        oeq, ogt, oge, olt, ole, one, ord, ueq, ugt, uge,
        ult, ule, une, uno,
        nuw, nsw, exact, inbounds,
        phi, call, select, shl, lshr, ashr, va_arg,
        trunc, zext, sext,
        fptrunc, fpext, fptoui, fptosi, uitofp, sitofp,
        ptrtoint, inttoptr, bitcast,
        ret, br, indirectbr, switch, invoke, unwind, unreachable,
        malloc, alloca, free, load, store, getelementptr,
        extractelement, insertelement, shufflevector,
        extractvalue, insertvalue,
    },
    alsoletter={\%},
    keywordsprefix={\%},% All identifiers starting with '%' will be printed as first order keywords.
    keywordstyle=[1]\bfseries,% As mentioned above, these are the keywords starting with '%', like '%5'
    keywordstyle=[2]\color{eminence}\bfseries,
    keywordstyle=[3]\color{darkgreen}\bfseries,
    keywordstyle=[4]\color{frenchplum},
}
\makeatother

\newtheorem{definition}{Definition}

\newtheorem{metric}{Metric}

\usepackage{tikz}
\makeatletter
\newenvironment{btHighlight}[1][]
{\begingroup\tikzset{bt@Highlight@par/.style={#1}}\begin{lrbox}{\@tempboxa}}
{\end{lrbox}\bt@HL@box[bt@Highlight@par]{\@tempboxa}\endgroup}

\newcommand\btHL[1][]{%
  \begin{btHighlight}[#1]\bgroup\aftergroup\bt@HL@endenv%
}
\def\bt@HL@endenv{%
  \end{btHighlight}%   
  \egroup
}
\newcommand{\bt@HL@box}[2][]{%
  \tikz[#1]{%
    \pgfpathrectangle{\pgfpoint{1pt}{0pt}}{\pgfpoint{\wd #2}{\ht #2}}%
    \pgfusepath{use as bounding box}%
    \node[anchor=base west, fill=orange!30,outer sep=0pt,inner xsep=1pt, inner ysep=0pt, rounded corners=3pt, minimum height=\ht\strutbox+1pt,#1]{\raisebox{1pt}{\strut}\strut\usebox{#2}};
  }%
}
\makeatother

\usepackage{booktabs}
\usepackage{xspace}
\newcommand{\ie}{\textit{i.e.,}\xspace}
\newcommand{\etal}{et al.\xspace}

\newcommand{\tool}{CROW\xspace}

\newcommand{\wasm}{WebAssembly\xspace}
\usepackage{relsize}% \smaller
\newcommand{\Rplus}{\protect\hspace{-.1em}\protect\raisebox{.22ex}{\smaller{\smaller\textbf{+}}}}
\newcommand{\CCpp}{\mbox{C/C\Rplus\Rplus}\xspace}
\newcommand{\blue}[1]{{\color{blue} #1 }}

\newcommand\nProgramsRosetta{\ensuremath{303}}
\newcommand\nProgramsRosettaDiversified{\ensuremath{239}}
\newcommand\nProgramsRosettaDiversifiedPercent{\ensuremath{79\%}}
\newcommand\nPreservedPercent{\ensuremath{99.48\%}}

\renewcommand{\blue}{}
\begin{document}

\renewcommand*{\sectionautorefname}{section}
\renewcommand*{\subsubsectionautorefname}{subsection}
\renewcommand*{\subsectionautorefname}{subsection}
\newcommand{\DTWStatic}{\ensuremath{\mathit{dt\_static}\xspace}}
\newcommand{\DTWDynamic}{\ensuremath{\mathit{dt\_dy}\xspace}}

\title{\tool: Code Diversification for \wasm}

\makeatletter
\newcommand{\linebreakand}{%
  \end{@IEEEauthorhalign}
  \hfill\mbox{}\par
  \mbox{}\hfill\begin{@IEEEauthorhalign}
}
\makeatother

 \author{\IEEEauthorblockN{Javier Cabrera Arteaga}
    \IEEEauthorblockA{\textit{KTH Royal Institute of Technology} \\
    javierca@kth.se}
    \and
    \IEEEauthorblockN{Orestis Floros}
    \IEEEauthorblockA{\textit{KTH Royal Institute of Technology} \\
    forestis@kth.se}
    \and
    \IEEEauthorblockN{Oscar Vera Perez}
    \IEEEauthorblockA{\textit{Univ Rennes, Inria, CNRS, IRISA} \\
    oscar.vera-perez@inria.fr}
    \linebreakand % <------------- \and with a line-break
    \IEEEauthorblockN{Benoit Baudry}
    \IEEEauthorblockA{\textit{KTH Royal Institute of Technology} \\
    baudry@kth.se}
    \and
    \IEEEauthorblockN{Martin Monperrus}
    \IEEEauthorblockA{\textit{KTH Royal Institute of Technology} \\
    martin.monperrus@csc.kth.se}
    
    }

\IEEEoverridecommandlockouts
\makeatletter\def\@IEEEpubidpullup{6.5\baselineskip}\makeatother
\IEEEpubid{\parbox{\columnwidth}{
    {\fontsize{7.5}{7.5}\selectfont Workshop on Measurements, Attacks, and Defenses for the Web (MADWeb) 2021 \\
    25 February 2021 \\
    ISBN 1-891562-67-3 \\
    https://dx.doi.org/10.14722/madweb.2021.23xxx \\
    www.ndss-symposium.org}
}
\hspace{\columnsep}\makebox[\columnwidth]{}}

\maketitle

\begin{abstract}
The adoption of \wasm increases rapidly, as it provides a fast and safe model for program execution in the browser. However, \wasm is not exempt from vulnerabilities that can be exploited by malicious observers.
Code diversification can mitigate some of these  attacks. In this paper, we present the first fully automated workflow for the diversification of \wasm binaries. We present \tool, an open-source tool implementing this workflow through enumerative synthesis of diverse code snippets expressed in the LLVM intermediate representation. We evaluate \tool's capabilities on \nProgramsRosetta{} C programs and study its use on a real-life security-sensitive program: libsodium, a modern cryptographic library. 
Overall, \tool is  able to  generate diverse variants  for \nProgramsRosettaDiversified{} out of $\nProgramsRosetta{}\,(\nProgramsRosettaDiversifiedPercent{})$ small programs.
Furthermore, our experiments show that our approach and tool is able to successfully diversify off-the-shelf cryptographic software (libsodium).
\end{abstract}

\section{Introduction}
% context
\wasm is the fourth official language of the Web \cite{WebAssemblyCoreSpecification}.
The language provides low-level constructs enabling efficient execution times, much closer to native code than JavaScript.
It constitutes a fast and safe platform to execute programs in the browser and embedded environments \cite{haas2017bringing}. Consequently, the adoption of \wasm has been rapidly growing since its introduction in 2015. Nowadays, languages such as Rust and \CCpp can be compiled to \wasm using mature toolchains and can be executed in all notable browsers.

% problem statement, goal
The \wasm execution model is designed to be secure and to prevent many memory and control flow attacks. Still, as its official documentation admits \cite{WebAssemblySecurity}, \wasm is not exempt from vulnerabilities that could be exploited \cite{usenixWASM2020}. Code diversification \cite{larsen2014sok,baudry2015multiple} is one additional protection that can harden the \wasm stack. 
This consists in synthesizing different variants of an original program that provide the same functionalities but exhibit different execution traces.
In this paper, we investigate the feasibility of diversifying \wasm code, which is, to the best of our knowledge, an unresearched area.

% our contribution
Our contribution is a workflow and a tool, called \tool, for automatic diversification of \wasm programs. It takes as input a \CCpp program and produces a set of diverse \wasm binaries as output. The workflow is based on enumerative code synthesis. 
First, \tool lists blocks that are potentially relevant for diversification,
second, \tool enumerates alternative instruction sequences,
and third, \tool checks that  the new instruction sequences are functionally equivalent to the original block.
\tool builds on the idea of superdiversification \cite{jacob2008superdiversifier} and extends the concept to the enumeration of a set of variants instead of synthesizing only one solution. We also take into account the specificities of \wasm and the details of its execution.

% evaluation
We evaluate the diversification capabilities of \tool in two ways. First, we diversify \nProgramsRosetta{} small C programs compiled to \wasm. Second, we run \tool to diversify a real-life cryptographic library that natively supports \wasm. In both cases, we measure  the diversity among binary code variants, as well as the diversity of execution traces. 
When measuring the diversity in binary code, we compare the \wasm and the machine code variants. This way we assess the ability of \tool at synthesizing variations in \wasm, as well as the extent to which these variations are preserved when compiling \wasm to machine code.
Our original experiments demonstrate the feasibility of diversifying \wasm code.
\tool generates diverse variants for $\nProgramsRosettaDiversified{}/\nProgramsRosetta{}\,(\nProgramsRosettaDiversifiedPercent{})$ C programs. TurboFan, the optimizing compiler used in the V8 engine, preserves \nPreservedPercent{} of these variants. \tool successfully synthesizes variants for the cryptographic library.
The variants indeed yield either different execution traces. This is promising milestone in getting a more secure Web environment through diversification.

To sum up, our contributions are:
\begin{itemize}
    \item \tool: the first automated workflow and tool to diversify \wasm programs, it generates many diverse \wasm binaries from a single input program.
    
    \item A quantitative evaluation over \nProgramsRosetta{} programs showing the capability of \tool to diversify \wasm binaries and measuring the impact of diversification on execution traces.

    \item A feasibility study of the diversification on a real-world \wasm program, demonstrating that \tool can handle libsodium, a state-of-the-art cryptographic library.
\end{itemize}

\section{Background}\label{sec:background}

\subsection{\wasm}
% What is \wasm
\wasm is a binary instruction format for a stack-based virtual machine.
It is designed to address the problem of safe, fast, portable and compact low-level code on the Web.
The language was first publicly announced in 2015 and since then, most major web browsers have implemented support for the standard.
Besides the Web, \wasm is independent of any specific hardware or languages and can run in a standalone Virtual Machine (VM) or in other environments such as Arduino~\cite{WARDuino2019}.
A paper by Haas \etal~\cite{haas2017bringing} formalizes the language and its type system, and explains the design rationale.

Listing \ref{CExample}  and \ref{WASMExample}  illustrate  \wasm.
Listing \ref{CExample} presents the C code of two functions and \autoref{WASMExample} shows the result of compiling these two functions into a \wasm module. The \lstinline[language=WAT]!type! directives at the top of the module declare the function: the types of its parameters and the type of the result.
Then, the definitions for the function follow. These definitions are sequences of stack machine instructions.
At the end, the \lstinline[language=WAT]!main! function is exported so that it can be called from outside this \wasm module, typically from JavaScript.
\wasm has four primitive types: integers (\lstinline[language=WAT]!i32! and \lstinline[language=WAT]!i64!) and floats (\lstinline[language=WAT]!f32! and \lstinline[language=WAT]!f64!) and it includes structured instructions such as
\lstinline[language=WAT]!block!,
\lstinline[language=WAT]!loop!
and \lstinline[language=WAT]!if!.

\begin{code}
\lstset{language=C,caption={C function that calculates the quantity $2x + x$},label=CExample,frame=b}
\begin{lstlisting}[style=CStyle]
int f(int x) { return 2 * x + x; }

int main(void) { return f(10); }
\end{lstlisting}
\lstset{
    language=WAT,
    caption={\wasm code  for \autoref{CExample}.},
    style=WATStyle,
    stepnumber=0,frame=b,
    label=WASMExample}
\begin{lstlisting}
(module
  (type (;0;) (func (param i32) (result i32)))
  (type (;1;) (func (result i32)))
  (func (;0;) (type 0) (param i32) (result i32)
    local.get 0
    local.get 0
    i32.const 2
    i32.mul
    i32.add)
  (func (;1;) (type 1) (result i32)
    i32.const 10
    call 0)
  (export "main" (func 1)))
\end{lstlisting}
\end{code}

% \wasm & security
\blue{
\wasm is characterized by an extensive security model~\cite{WebAssemblySecurity} founded on a sandboxed execution environment that provides protection against common security issues such as data corruption, code injection and return oriented programming (ROP). However, WebAssembly is no silver bullet and is vulnerable under certain conditions  \cite{usenixWASM2020}.  This motivates our work on software diversification as one possible mitigation among the wide range of security counter-measures.}

\subsection{Motivation for Moving Target Defense in the Web}

The distribution model for web computing is as follows: build one binary and distribute millions of copies, all over the world, which run on browsers. In this model an attacker has two key advantages over the developers: she has a runtime environment that she fully controls and observes in any possible way. Consequently, when she finds a flaw in this virtually transparent environment, knowing that this flaw is present in the millions of copies that have been distributed over the world, she can exploit the flaw at scale.

The developers can never assume that they can control the web browser. Yet, they can challenge the second advantage of the attacker, known as the break-once-break-everywhere advantage. The developers can stop distributing clones of the binary and distribute diverse versions instead, as suggested by the pioneering software diversification works of Cohen~\cite{cohen1993operating} and Forrest \etal~\cite{forrest1997building}.

\blue{In the context of diversification, moving target defense~\cite{taguinod2015toward}  means distributing diverse variants constantly. In the context of the web, it means distributing a different variant at each HTTP request.
Moving target defense is appropriate for mitigating yet unknown vulnerabilities. The diversification technique does not always remove the potential flaws, yet the vulnerabilities in the diversified binaries can be located in different places.
With moving target defense, a successful attack on one browser cannot be performed on another browser with the same effectiveness. The diversified binaries that \tool outputs can be used interchangeably over the network, in a moving target defence choreographed over the web. 
}%

\blue{%
To sum up, by combining moving target defense deployment to diversification, we reduce the information asymmetry between the Web attacker and the defender, increasing the uncertainty and complexity of successful attacks over all client browsers \cite{cui2011symbiotes,zhuang2014towards}.
}

\section{\tool's Diversification Technique}\label{workflow:algorithm}
In this section we describe the workflow of \tool for diversifying \wasm programs. First we introduce the main concepts behind \tool. Then, we describe each stage of the workflow and we discuss the key implementation details.

\subsection{Definitions}
In this subsection we define the key concepts for \tool. 

\begin{definition}{Block (based on Aho \etal \cite{10.5555/6448}):}\label{def:code-block}
    Let $P$ be a program. A block $B$ is a grouping of declarations and statements in $P$ inside a function $F$. 
\end{definition}

\begin{definition}{Program state (based on Mangpo \etal \cite{10.1145/2980024.2872387}):}\label{def:program-state}
    At any point in time, the program state $S$ is defined as the collection of local and global variables, and, the program counter pointing to the next instruction.
\end{definition}

\begin{definition}{Pure block:}\label{def:pure-code}
    A block $B$ is said to be pure if and only if, given the program state $S_i$, every execution of $B$ produces the same state $S_o$.
\end{definition}

\begin{definition}{Functional equivalence modulo program state (based on Le \etal \cite{10.1145/2594291.2594334}):}\label{def:functional-equivalence}
    Let $B_1$ and $B_2$ be two blocks. We consider the program state before the execution of the block, $S_i$, as the input and the program state after the execution of the block, $S_o$, as the output. $B_1$ and $B_2$ are functionally equivalent if given the same input $S_i$ both codes produce the same output $S_o$.
\end{definition}
 
\begin{definition}{Code replacement:}\label{def:code-replacement}
    Let $P$ be a program and $T$ a pair of blocks $(B_1, B_2)$. $T$ is a candidate code replacement if $B_1$ and $B_2$ are both pure as defined in \autoref{def:pure-code} and functionally equivalent as defined in \autoref{def:functional-equivalence}.
    Applying $T$ to $P$ means replacing $B_1$ by $B_2$. The application of $T$ to $P$ produces a program variant $P'$ which consequently is functionally equivalent to $P$.     
\end{definition}

\tool generates new program variants by finding and applying code replacements as defined in \autoref{def:code-replacement}. 
A program variant could be produced by applying more than one candidate code replacement. 
For example, the tuple, composed by the code blocks in \autoref{example:block1} and \autoref{example:block2}, is a code replacement for \autoref{WASMExample}.

\lstdefinestyle{ccode}{
  numbers=left,
  stepnumber=1,
  numbersep=10pt,
  tabsize=4,
  showspaces=false,
  showstringspaces=true,
    moredelim=**[is][{\btHL[fill=black!10]}]{`}{`},
    moredelim=**[is][{\btHL[fill=celadon!40]}]{@}{@}
}

\lstdefinestyle{nccode}{
  numbers=none,
  stepnumber=1,
  numbersep=10pt,
  tabsize=4,
  showspaces=false,
  breaklines=true, 
  showstringspaces=false,
    moredelim=**[is][{\btHL[fill=black!10]}]{`}{`},
    moredelim=**[is][{\btHL[fill=celadon!40]}]{@}{@}
}

\lstdefinestyle{watcode}{
  numbers=none,
  stepnumber=1,
  numbersep=10pt,
  tabsize=4,
  showspaces=false,
  breaklines=true, 
  showstringspaces=false,
    moredelim=**[is][{\btHL[fill=black!10]}]{`}{`},
    moredelim=**[is][{\btHL[fill=celadon!40]}]{!}{!}
}

{\captionsetup{width=0.45\linewidth}
\noindent\begin{minipage}[b]{0.45\linewidth}
    \lstset{
        language=WAT,
        style=watcode,
        basicstyle=\footnotesize\ttfamily,
        columns=fullflexible,
        breaklines=true}
        \begin{lstlisting}[label=example:block1,caption={\wasm pure code block from \autoref{WASMExample}.},frame=b]{Name}
  `local.get 0`
  `i32.const 2`
  `i32.mul`            ; 2 * x ;
        \end{lstlisting}
   \end{minipage}\hfill%
\noindent\begin{minipage}[b]{0.45\linewidth}
    \lstset{
        language=WAT,
                        style=watcode,
        basicstyle=\footnotesize\ttfamily,
                        columns=fullflexible,
                        breaklines=true}
        
        \begin{lstlisting}[label=example:block2,caption={Code block that is functionally equivalent to \autoref{example:block1} },frame=b]{Name}
  !local.get 0!
  !i32.const	1!
  !i32.shl!  ; x << 1 ;
        \end{lstlisting}
\end{minipage}
}

\label{sec:overview}
\subsection{Overview}
\tool synthesizes variants for \wasm programs. We assume that the programs are generated through the LLVM compilation pipeline. This assumption is motivated as follows: first, LLVM-based compilers are the most popular compilers to build \wasm programs \cite{usenixWASM2020}; second, the availability of source code (typically \CCpp for \wasm) provides a structure to perform code analysis and produce code replacements that is richer than the binary code.

\tool takes as input a \CCpp program and produces a set of unique, diversified \wasm binaries.
Figure \ref{diagrams:flow} shows the stages of this workflow. The workflow starts with compiling the input program into LLVM bitcode using clang. Then, \tool analyzes the bitcode to identify all pure blocks and to synthesize a set of candidate replacements for each pure block. This is what we call the \emph{exploration} stage. 
In the \emph{generation} stage, \tool combines the candidate code replacements to generate different LLVM bitcode variants. Finally, those bitcode variants are compiled to \wasm binaries that can be sent to web browsers.

\emph{Challenges.} The concept of diversifying \wasm programs is novel and it is arguably hard for the following reasons. First, \wasm is a structured binary format, without goto-like instructions. This  prevents the direct application of a wide range of diversification operators based on goto~\cite{wang01}.
Second, the existing transformation and diversification tools target instruction sets larger than the one of \wasm \cite{seibert2014information}. This limits the efficiency of diversification, and the possibility of searching for a large number of equivalent code replacements.
We address the former challenge using the LLVM intermediate representation as the target for diversification. We address the latter challenge  by tailoring a  superoptimizer for  LLVM, using its subset of the LLVM intermediate representation. In particular, we prevent the superoptimizer from synthesizing  instructions that have no correspondence in \wasm (for example, \texttt{freeze} instructions), which is an essential step to get executable diversified \wasm code.

\begin{figure*}
  \includegraphics[width=\linewidth]{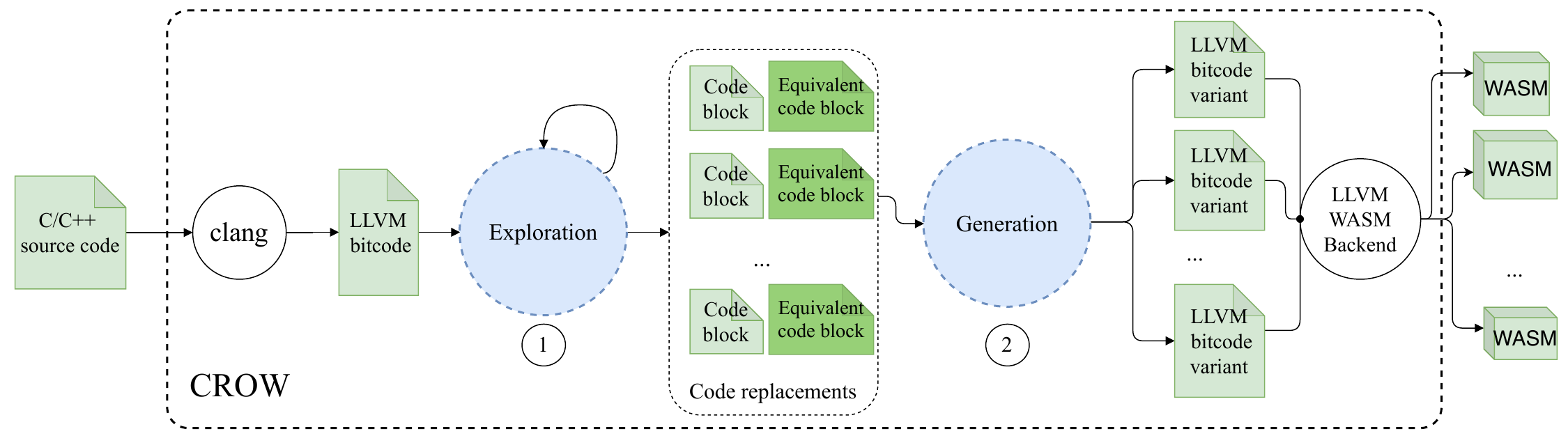}
  \caption{\tool's workflow for diversifying \wasm programs.}
  \label{diagrams:flow}
\end{figure*}

\subsection{Exploration stage}
\label{workflow:exploration}
Given a program $P$ for which we want to generate \wasm variants, the exploration stage of \tool identifies all pure blocks in the LLVM bitcode of $P$. \tool considers every directed acyclic graph contained in one function as a pure block. Then, \tool searches for code replacements for each one of them. 

The generation of a code replacement consists of two steps. First, the synthesis of the new block, and, second, equivalence checking. Every variant block that passes the equivalence check is stored for use in diversification.
The synthesis of block variants consists of enumerating all possible blocks that can be built as a combination of a given number of instructions, bounded by a maximum value to keep a tractable synthesis space. 

% discussion on performance
There are two parameters to control the size of the search space and hence the time required to traverse it.
On one hand, one can limit the size of the variants. In our experiments we limit the block variants to a maximum of $50$ instructions. On the other hand, one can limit the set of instructions that are used for the synthesis. In our experiments, we use between $1$ instruction (only additions) and $60$ instructions (all supported instructions in the synthesizer).
This configuration allows the user to find a trade-off between the amount of variants that are synthesized and the time taken to produce them. 

% commented out to save space
\lstdefinelanguage{LLVM}
    {morekeywords={i32,mul,align,nsw,add,load,store,define,br, ret, shl, ret},
    sensitive=false,
    morecomment=[l]{;},
    morecomment=[s]{;}{;},
    morestring=[b]",
}
\lstdefinestyle{nccode}{
    numbers=left,
    tabsize=4,
    showspaces=false,
    breaklines=true, 
    showstringspaces=false,
    moredelim=**[is][{\btHL[fill=black!10]}]{`}{`},
    moredelim=**[is][{\btHL[fill=celadon!40]}]{!}{!}
}
\lstset{
    language=LLVM,
    style=nccode,
    %basicstyle=\small\ttfamily,
    columns=fullflexible,
    breaklines=true
}

\begin{code}
\centering
\captionof{lstlisting}{\autoref{CExample} in LLVM's intermediate representation.}\label{example:original:llvm}
\lstset{numbers=none}
\noindent\begin{minipage}[b]{.55\linewidth}
\centering
\begin{lstlisting}[xleftmargin=1em]
define i32 @f(i32) {
 %2 = mul nsw i32 %0,2 %*\label{line:original:first:start}*)
 %3 = add nsw i32 %0,%2 %*\label{line:original:second:start}*)
 
 ret i32 %3
}

define i32 @main() {
 %1 = tail call i32 @f(i32 10)
 ret i32 %1
}
\end{lstlisting}
\end{minipage}\hfill%
\noindent\begin{minipage}[b]{.4\linewidth}
\vspace{-10pt}
    \lstdefinestyle{nccode}{
      tabsize=4, 
      showspaces=false,
      breaklines=true, 
      showstringspaces=false,
    moredelim=**[is][{\btHL[fill=black!10]}]{`}{`},
    moredelim=**[is][{\btHL[fill=celadon!40]}]{!}{!}
    }
    \lstset{
        language=LLVM,
        style=nccode,
        columns=fullflexible,
        breaklines=true,
        belowcaptionskip=1pt,
        abovecaptionskip=1pt,
    }
    \vfill%
    \begin{lstlisting}[label={ref:block1}, caption={Block A},title={Block A},name={A}]
`%2 = mul nsw i32 %0,2`
    \end{lstlisting}
    \begin{lstlisting}[caption={Block B}, title={Block B},label={ref:block2},name={B}]
!%2 = mul nsw i32 %0,2!
!%3 = add nsw i32 %0,%2!
    \end{lstlisting}
\end{minipage}
\end{code}
In \autoref{example:original:llvm} we illustrate the LLVM bitcode representation of \autoref{CExample}. In this bitcode, \tool identifies two pure blocks in function \texttt{f()}, which are displayed on the right part of the listing, in gray and green.  The first pure block is composed of one single instruction (line \ref{line:original:first:start}) that performs the \lstinline{2*x} multiplication. The second block has two instructions, one multiplication and one addition.

Using \tool, it is possible to diversify both blocks.
For example, using a maximum of $1$ instruction per replacement and searching over the complete bitcode instruction set, a potential replacement for Block A is:
\texttt{\%2 = shl nsw i32 \%0,1 \%}.
This replacement calculates the same expression \texttt{2*x}, using a shift left operation.

\blue{To determine the equivalence between a pure block and a candidate replacement, we use an equivalence checker based on SMT \cite{2008_Moura_SMT}. In our example, the checker would prove that there cannot be a value of $x$ such that $2*x \neq x \ll 1$. In general, if no such counter-example exists, then the functional equivalence is assumed. On the other hand, if there exists an input resulting in different outputs for a block and a variant, then they are proven not equivalent and the variant is discarded.}

\subsection{Generation stage}
\label{workflow:generation}
In this stage, we select and combine code replacements that have been synthesized during the exploration stage, in order to generate \wasm binary variants.
We apply each code replacement to the original program to produce a LLVM IR variant.
Then, this IR is compiled into a \wasm binary. 
\tool generates \wasm binaries from all possible combinations of code replacements as the power set over all code replacements. 

After the exploration phase, it is possible that two subsets of code replacements overlap, \ie they produce the same \wasm binary. The overlap between blocks is explained as follows: Let $S = \{(B_1, R_1), (B_1, R_2), \cdots, (B_n, R_m)\}$ be a set of candidate replacements over a program $P$. If two blocks from the original program $B_i$, $B_j$, $j \neq i$, overlap, \ie the intersection of $\CFG(B_i)$\footnote{$\CFG(A)$ refers to backward Control Flow Graph starting at inst. $A$.} and $\CFG(B_j)$ is not empty, then only the replacements of the largest original block are preserved when combining blocks.

In this example, the exploration stage synthesizes $6 + 1$ bitcode variants for the considered blocks respectively, which results in $14$ module variants (the power set combination size). Yet, the generation stage would eventually generate $7$ variants from the original \wasm binary. This gap between the number of potential and the actual number of variants is a consequence of the redundancy among the bitcode variants when composing several variants into one. 

\subsection{Implementation}
\label{workflow:implementation}

% Souper general architecture and concepts
\blue{The majority of the \wasm applications are built from C/C++ source code using the LLVM toolchain. 
Consequently, the implementation of \tool is based on LVVM/
Furthermore, \tool extends Souper~\cite{Sasnauskas2017Souper:Superoptimizer}, a superoptimizer for LLVM that aims to reduce the size of binary code.
Souper has its own intermediate representation, which is a subset of the LLVM IR.
}

\blue{To extract code blocks, we scan LLVM modules, looking for instructions that return integer-typed values. Each such instruction is considered as the exit of a code block. Souper's representation of a code block is built as a backward traversal process through the dependencies of the detected instruction.  If memory loads or function calls are found, the backward traversal  process is stopped and the current instruction is considered as an input variable for the code block. Notice that, by construction, Souper's translation is oblivious to the memory model, thus, it cannot infer string data types or other abstract data types. The translation from Souper IR to a BitVector SMT theory is done on the fly. Souper uses the z3\footnote{https://github.com/Z3Prover/z3} solver to check the equivalence between a code block original and a potential replacement for it.}

We now summarize the main changes that we implement in Souper and in the LLVM backend in order to support diversification.
Souper, as a superoptimizer, aims at generating a single variant that is smaller than the original, yet we want to obtain as many blocks as possible. To achieve automatic diversification, we modify Souper to disable the key cost restriction functions, data-flow pruning and peephole optimizations, all being detrimental for diversification. \blue{In order to increase the number of variants that \tool can generate, \tool parallelizes the process of replacement synthesis.}

In addition, \tool orchestrates a series of Souper executions with various configurations (in particular the size of the replaced expression). Finally, we carefully fine-tune a set of $19$ Souper options to ensure that the search is effective for diversification in feasible time. 

In the generation stage of \tool, we also modify Souper to amplify the generation of \wasm binary diversity. Initially, Souper generates a single bitcode variant, inserting all replacements at once. We modify it so that we can obtain a combination of code replacements. 
% peephole optimizations
\blue{Finally, on the LLVM side, we disable all peephole optimizations in the \wasm backend, in particular instructions merging and constant folding}. This aims to preserve the variations introduced in the LLVM bitcode during the generation of binaries.

The implementation of \tool is publicly available for sake of open science and can be reviewed at \url{https://github.com/KTH/slumps/tree/master/crow}.

%%%%%%%%%%%%%%%%%%%%%%%%%%%%%%%%%%%%%%%%%%%%%%%%%%%
%%%%%%%%%%%%%%%%%%%%%%%%%%%%%%%%%%%%%%%%%%%%%%%%%%%
%%%%%%%%%%%%%%%%%%%%%%%%%%%%%%%%%%%%%%%%%%%%%%%%%%%

\section{Evaluation Protocol}\label{sec:evaluation}
\newcommand\rqstatic{To what extent are the program variants generated by \tool statically different?\xspace}
\newcommand\rqstaticmachine{To what extent the generated variant replacements are maintained during machine code translation?\xspace}
\newcommand\rqdynamic{To what extent are the program variants generated by \tool dynamically different?\xspace}
\newcommand\rqperformance{To what extent are the execution times of program variants generated by \tool different?}
\newcommand\rqlibsodium{To what extent can \tool be applied to diversify real-world security-sensitive software?\xspace}

To evaluate the capabilities of \tool to diversify \wasm programs, we formulate the following research questions:
{%
% XXX: About the use of enumerate for this purpose:
% https://github.com/programming-journal/programming/issues/39
\setlength\leftmargini{2.5em}% Seems to be the original default. Fixes misalignment with custom label.
\begin{enumerate}[label=RQ\arabic*:, ref=RQ\arabic*]
    \item \label{rq:static} \textbf{\rqstatic}
    We  check whether the \wasm binary variants produced by \tool are different from the original \wasm binary. Then, we assess whether the generation of x86 machine code performed by  V8's \wasm engine preserves \tool's transformations.
    
    \item \label{rq:dynamic}\textbf{\rqdynamic} 
    It is known that not all diversified programs produce distinguishable executions \cite{crane2015thwarting}, sometimes it is impossible to observe different behaviors between variants. We check for the presence of different behaviors with a custom \wasm interpreter, characterizing the behavior of a \wasm program by its stack operation trace.
    
    \item \textbf{\rqlibsodium}
    We assess the ability of \tool to diversify a state-of-the-art cryptographic library for \wasm, libsodium~\cite{libsodium}.
\end{enumerate}%
}

\subsection{Corpus}
We answer \ref{rq:static} and \ref{rq:dynamic} with a corpus of programs appropriate for our experiments. We take programs from the  Rosetta Code project\footnote{\url{http://www.rosettacode.org/wiki/Rosetta_Code}}. This website hosts a curated set of solutions for specific programming tasks in various  programming languages.
It contains a wide range of tasks, from simple ones, such as adding two numbers, to complex algorithms like a compiler lexer. We first collect all C programs from Rosetta Code, which represents $989$ programs as of 01/26/2020. 
Next, we apply a number of filters.
We discard
\begin{enumerate*}
\item all programs that do not compile with \texttt{clang},
\item all interactive programs requiring input from users \ie invoking functions like \texttt{scanf},
\item all programs that contain more than 100 blocks,
\item all programs without termination,
\item all programs with non-deterministic operations, for example, programs working with time or random functions. This filter produces a final set of \nProgramsRosetta{} programs
\end{enumerate*}.

The result is a corpus of \nProgramsRosetta{} C programs. These programs range from $7$ to $150$ lines of code and solve a variety of problems, from the \textit{Babbage} problem to  \textit{Convex Hull} calculation.  

\subsection{Protocol for RQ1}

With \ref{rq:static},
we assess the ability of \tool to generate \wasm binaries that are different from the original program.
For this, we compute a distance metric between the original \wasm binary and each binary generated by \tool. 
Since \wasm binaries are further transformed into machine code before they execute, we also check that this additional transformation preserves the difference introduces by \tool in the \wasm binary. We use the Turbofan ahead-of-time compiler of V8, with all its possible optimizations, to generate a x86 binary for each \wasm binary. Then, we compare the x86 version of each variant against the x86 binary corresponding to the original \wasm binary.

We compare the \wasm and machine code of each program and its variant using Dynamic Time Warping (DTW)  \cite{Maia08usinga}. DTW computes the global alignment between two sequences. It returns a value capturing the cost of this alignment, which is actually a distance metric, called DTW. The larger the DTW distance, the more different the two sequences are. In our case, we compare the sequence of instructions of each variant with the initial program and the other variants. We obtain two DTW distance values for each program-variant pair: one at the level of \wasm code and the another one at the level of x86 code. \autoref{metric:static1} below defines these metrics.

\begin{metric}{\DTWStatic:}\label{metric:static1}
Given two programs $P_X$ and $V_X$ written in $X$ code, \DTWStatic{}($P_X$, $V_X$), computes the DTW distance between the corresponding program instructions for representation $X$ ($X \in \{Wasm, x86\}$). A \DTWStatic{}($P_X$, $V_X$) of $0$ means that the code of both the original program and the variant  is the same, i.e., they are statically identical in the representation $X$. The higher the value of \DTWStatic{}, the more different the programs are in representation X. \\
\end{metric}

% When comparing machine code programs, we remove the code generated to support the WASI functions. This code is glued to the compiled \wasm, it is always the same at any representation and it is not targeted by \tool.

We run \tool on our corpus of \nProgramsRosetta{} programs. We configure \tool to run with a diversification timeout of 6 hours per program. For each program, we collect the set of generated variants. For all pairs program,  variant that are different, we compute both \DTWStatic{} for \wasm and x86 representations. 

The key property we consider is as follows: if \DTWStatic{}($P_{Wasm}$, $P_{Wasm}'$) $>$ 0 and \DTWStatic{}($P_{x86}$, $P_{x86}'$) $>$ 0, this means that  both programs are still different when compiled to machine code, and we conclude that V8's compiler does not remove the transformations made by \tool.  Notice that, this property only makes sense between variants of the same program (including the original).

\subsection{Protocol for RQ2}\label{sec:protocol-rq2}
For \ref{rq:dynamic}, we compare  the executions of a program and its variants for a given input. In this experiment, we characterize the  execution of a \wasm binary according to its trace of stack operations.

This method of tracing allows us to evaluate \tool's effect on program execution according to the \wasm specification, independently of any specific engine.

For each execution of a \wasm program, we collect a trace of stack operations. These traces are composed of stack-type instructions: \texttt{push <value>} and \texttt{pop <value>}. 
All traces are ordered with respect to the timestamp of the events. We compare the traces of the original program against those of the variants with DTW. DTW computes the global alignment between two traces and provides a value for the cost of this alignment.

\begin{metric}{\DTW{}:}\label{metric:stack}
Given a program P and a \tool generated variant P', \DTW{}(P,P'), computes the DTW distance between the corresponding stack operation traces collected during their execution. A \DTW{} of $0$ means that both traces are identical. The higher the value, the more different the stack operation traces. 
\end{metric}

To answer \ref{rq:dynamic} we compute \autoref{metric:stack} for a study subject program and all the unique program variants generated by \tool in a pairwise comparison. 
The pairwise comparison allows us to compare the diversity between variants as well.
We use SWAM\footnote{\url{https://github.com/satabin/swam}} to collect the stack operation traces. SWAM is a \wasm interpreter that provides functionalities to capture the dynamic information of \wasm program executions including the stack operations. We compute the DTW distances with STRAC~\cite{Cabrera19}. 

The builtin WebAssembly API for JavaScript is usually mutable, thus, the same model for traces collection can be implemented on top of V8. In other words, a custom interpreter can be implemented in order to collect the traces in the browser or standalone JavaScript engines. This validates the usage of SWAM to study the traces diversity.

\subsection{Protocol for RQ3}\label{sec:protocol-rqlibsodium}
% Intro to RQ3
In RQ3, we assess the ability of \tool to diversify a mature and complex software library related to security. 
% Why libsodium?
We choose the libsodium~\cite{libsodium} cryptographic library, which natively compiles to \wasm.
With $3752$ commits contributed by $96$ developers, its API provides the basic blocks for encryption, decryption, signatures and password hashing.
We experiment with code revision
\href{https://github.com/jedisct1/libsodium/tree/2b5f8f2b6810121c2d9a8cc8a392e01f4d3de433}{2b5f8f2b}, which
contains $45232$ lines of C code.
Libsodium has $102$ separate \wasm modules that we use as input for \tool.
Each module corresponds to one C file that encompasses a set of related functions.

To answer RQ3, we run \tool on the libsodium bitcodes, generating a set of \wasm variants. 
Then, we assess both binary code diversity and behavioural diversity between the variants and the original libsodium, using the same techniques as in RQ1 and RQ2.

%\emph{Toolchain preparation}. We need to adapt the build chain of libsodium for \tool. First, we wrap the default compiler (clang) used by libsodium in order to be able to save the LLVM intermediate bitcode for each source code file. These bitcodes are fed to \tool, which generates variants. Then, we replace some of the original intermediate non-diversified bitcodes with the variants produced by \tool and resume the compilation. The final result is a libsodium \wasm module that combines both original and diversified compilation objects.

\emph{Collecting traces}
% explaining the composition of variants and test suite
The libsodium repository includes an extensive test suite of $77$ tests, where one test is one usage scenario.
We use this test suite to measure the trace diversity among program variants. 
Since some test traces are larger than $1$ GB each, we focus on reasonably sized tests: we select the $41/77$ test cases that produce a trace containing less than $50$ million events each.

To measure the relative trace diversification for each test, we normalize the \DTW{} used in RQ2 by dividing it with the length of the original trace.
This allows us to compare the relative success of \tool's diversification technique across different tests.

\blue{%
Since libsodium uses a pseudo-number generator, we set a static seed when executing libsodium, so that the diversity observed in traces is only due to \tool's diversification.
In \wasm, libsodium generates random numbers using the ChaCha20~\cite{bernstein2012high} cipher through the \texttt{arc4random} API.
To quantify the effectiveness of our diversification technique, we compare the trace distance produced by our technique with the trace distance that occurs when the seed is changed (baseline).
}

\section{Experimental Results}\label{sec:results}
In this section we present the results for the research questions formulated in \autoref{sec:evaluation}.

\begin{figure}
    \centering
      \includegraphics[width=\linewidth]{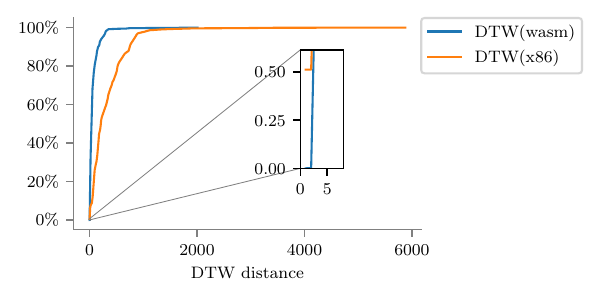}
      \caption{Cumulative distribution for all pairwise comparisons between a program and its variants. Each line corresponds to a different program representation.}
      \label{fig:rq1}
\end{figure}

\subsection{\rqstatic}\label{results:rq1}

We run \tool on  \nProgramsRosetta{} C programs compiled to \wasm. \tool produces at least one unique program variant for $\nProgramsRosettaDiversified{}/\nProgramsRosetta{}$ programs. For the rest of the programs ($64/\nProgramsRosetta{}$), the timeout is reached before \tool can find any valid variant. 

We subsequently perform a  manual analysis of the programs that yield more than 100 unique \wasm variants.
This reveals one key reason that favors a large number of unique \wasm variants: the programs include  bounded loops. In these cases \tool
synthesizes variants for the loops by unrolling them. Every time a loop is unrolled, the loop body is copied and moved as part of the outer scope of the loop. This creates a new, statically different, program. The number of programs grows exponentially with nested loops. 

A second key factor for the synthesis of many variants relates to the presence of arithmetic. Souper, the synthesis engine used by \tool, is effective in replacing  arithmetic instructions by equivalent instructions that lead to the same result. For example, \tool generates unique variants by replacing multiplications with additions or shift left instructions (\autoref{add:example}). Also, logical comparisons are replaced, inverting the operation and the operands (\autoref{cmp:examples}). 

\lstdefinelanguage{LLVM}
    {morekeywords={i32,mul,align,nsw,add,load,store,define,br, ret, shl, ret},
    sensitive=false,
    morecomment=[l]{;},
    morecomment=[s]{;}{;},
    morestring=[b]",
}
\lstdefinestyle{nccode}{
    numbers=none,
    firstnumber=1,
    stepnumber=1,
    numbersep=10pt,
    tabsize=4,
    showspaces=false,
    breaklines=true, 
    showstringspaces=false,
    moredelim=**[is][{\btHL[fill=black!10]}]{`}{`},
    moredelim=**[is][{\btHL[fill=celadon!40]}]{!}{!}
}
\lstset{
    language=WAT,
    style=nccode,
    basicstyle=\footnotesize\ttfamily,
    columns=fullflexible,
    breaklines=true
}
\begin{code}
\noindent\begin{minipage}[b]{0.48\linewidth}
    \captionof{lstlisting}{Diversification through arithmetic expression replacement.}\label{add:example}
    \noindent\rule{\linewidth}{0.4pt}
    \noindent\begin{minipage}[t]{0.46\linewidth}
        \begin{lstlisting}
local.get 0
`i32.const 2`
`i32.mul`
        \end{lstlisting}
    \end{minipage}%
    \hfill\noindent\begin{minipage}[t]{0.46\linewidth}
        \lstdefinestyle{nccode}{
            numbers=none,
            firstnumber=2,
            stepnumber=1,
            numbersep=10pt,
            tabsize=4, 
            showspaces=false,
            breaklines=true, 
            showstringspaces=false,
            moredelim=**[is][\btHL]{`}{`},
            moredelim=**[is][{\btHL[fill=black!10]}]{`}{`},
            moredelim=**[is][{\btHL[fill=celadon!40]}]{!}{!}
        }
        \lstset{
            language=WAT,
            style=nccode,
            basicstyle=\footnotesize\ttfamily,
            columns=fullflexible,
            breaklines=true
        }
        \begin{lstlisting}
local.get 0
!i32.const 1!
!i32.shl!
        \end{lstlisting}
    \end{minipage}
\end{minipage}\hfill%
\begin{minipage}[b]{0.48\linewidth}
    \captionof{lstlisting}{Diversification through inversion of comparison operations.}\label{cmp:examples}
    \noindent\rule{\linewidth}{0.4pt}
    \begin{minipage}[t]{.46\linewidth}
        \begin{lstlisting}
`local.get 0`
`i32.const 10`
`i32.gt_s`
        \end{lstlisting}
    \end{minipage}\hfill\begin{minipage}[t]{.46\linewidth}
        \lstdefinestyle{nccode}{
            numbers=none,
            firstnumber=2,
            stepnumber=1,
            numbersep=10pt,
            tabsize=4, 
            showspaces=false,
            breaklines=true, 
            showstringspaces=false,
            moredelim=**[is][{\btHL[fill=black!10]}]{`}{`},
            moredelim=**[is][{\btHL[fill=celadon!40]}]{!}{!}
        }
        \lstset{
            language=WAT,
            style=nccode,
            basicstyle=\footnotesize\ttfamily,
            columns=fullflexible,
            breaklines=true
        }
        \begin{lstlisting}
!i32.const 11!
!local.get 0!
!i32.le_s!
        \end{lstlisting}
    \end{minipage}%
\end{minipage}
\end{code}

\lstdefinelanguage{LLVM}
    {morekeywords={i32,mul,align,nsw,add,load,store,define,br, ret, shl, ret},
    sensitive=false,
    morecomment=[l]{;},
    morecomment=[s]{;}{;},
    morestring=[b]",
}
\lstdefinestyle{nccode}{
    numbers=none,
    firstnumber=1,
    stepnumber=1,
    numbersep=10pt,
    tabsize=4,
    showspaces=false,
    breaklines=true, 
    showstringspaces=false,
    moredelim=**[is][{\btHL[fill=black!10]}]{`}{`},
    moredelim=**[is][{\btHL[fill=red!10]}]{!}{!}
}
\lstset{
    language=WAT,
    style=nccode,
    basicstyle=\footnotesize\ttfamily,
    columns=fullflexible,
    breaklines=true
}
\begin{code}
\centering
\captionof{lstlisting}{Excerpt of \wasm program p74: CROW replaces a loop by a constant.}\label{babbage_code}
\noindent\rule{\linewidth}{0.4pt}
\noindent\begin{minipage}[t]{.45\linewidth}
    \begin{lstlisting}
local.set 1
`loop  ;; label = @1`
  ...
end
...
i32.store
    \end{lstlisting}
\end{minipage}\hfill%
\noindent\begin{minipage}[t]{.45\linewidth}
    \lstdefinestyle{nccode}{
      numbers=none,
      firstnumber=2,
      stepnumber=1,
      numbersep=10pt,
      tabsize=4, 
      showspaces=false,
      breaklines=true, 
      showstringspaces=false,
    moredelim=**[is][{\btHL[fill=black!10]}]{`}{`},
    moredelim=**[is][{\btHL[fill=celadon!40]}]{!}{!}
    }
    \lstset{
        language=WAT,
        style=nccode,
        basicstyle=\footnotesize\ttfamily,
        columns=fullflexible,
        breaklines=true
    }
    \begin{lstlisting}

local.get 0
!i32.const 25264!


i32.store
    \end{lstlisting}
\end{minipage}
\end{code}

We now discuss the prevalence of the transformations made by CROW when the \wasm binaries are transformed to machine code, specifically with the V8's engine. In \autoref{fig:rq1} we plot the cumulative distribution of
\DTWStatic{}, comparing \wasm binaries (in blue) and x86 binaries (in orange). The figure plots  a total of 103003 \DTWStatic{} values for each representation, two values for each variant pair comparison (including original) for the 239 program.
The value on the y-axis shows which percentage of the total comparisons lie below the corresponding \DTWStatic{} value on the x-axis.
Since we measure the distances between original programs and \wasm variants, then $100\%$ of these  binaries have $\DTWStatic{}>0$.
Let us consider the x86 variants: \DTWStatic{} is strictly positive for \nPreservedPercent{} of variants. In all these cases, the V8 compilation phase does not undo the \tool diversification transformations.
Also, we see that there is a gap between both distributions, the main reason is the natural inflation of machine code. For example, two variants that differ by one single instruction in \wasm, can be translated to machine code where the difference is increased by more than one machine code instruction.

% Negative prevalence
The zoomed subplot focuses on the beginning of the distribution, it shows that the \DTWStatic{} is zero for $0.52\%$ of the x86 binaries.
In these cases the V8 TurboFan compiler from \wasm to x86 reverts the \tool transformations.
We find that \tool produces at least one of these reversible transformations for $34/\nProgramsRosettaDiversified{}$ programs.
\autoref{add:prevalence_example} shows one of the most common transformations that is reversed by TurboFan, according to our experiments.

%Besides, local variables reordering and common subexpressions are cases that TurboFan reverses.
\lstdefinelanguage{LLVM}
    {morekeywords={i32,mul,align,nsw,add,load,store,define,br, ret, shl, ret},
    sensitive=false,
    morecomment=[l]{;},
    morecomment=[s]{;}{;},
    morestring=[b]",
}
\lstdefinestyle{nccode}{
    numbers=none,
    firstnumber=1,
    stepnumber=1,
    numbersep=10pt,
    tabsize=4,
    showspaces=false,
    breaklines=true, 
    showstringspaces=false,
    moredelim=**[is][{\btHL[fill=black!10]}]{`}{`},
    moredelim=**[is][{\btHL[fill=celadon!40]}]{!}{!}
}
\lstset{
    language=WAT,
    style=nccode,
    basicstyle=\footnotesize\ttfamily,
    columns=fullflexible,
    breaklines=true
}
\begin{code}
\noindent\begin{minipage}[b]{0.9\linewidth}
    \captionof{lstlisting}{Replacement in \wasm that is translated to the same x86 code by V8-TurboFan.}\label{add:prevalence_example}
    \noindent\rule{\linewidth}{0.4pt}
    \begin{minipage}[t]{0.45\linewidth}
        \begin{lstlisting}
`i32.const -<n>`
`i32.sub`
        \end{lstlisting}
    \end{minipage}%
    \hfill\noindent\begin{minipage}[t]{0.45\linewidth}
        \lstdefinestyle{nccode}{
            numbers=none,
            firstnumber=2,
            stepnumber=1,
            numbersep=10pt,
            tabsize=4, 
            showspaces=false,
            breaklines=true, 
            showstringspaces=false,
            moredelim=**[is][\btHL]{`}{`},
            moredelim=**[is][{\btHL[fill=black!10]}]{`}{`},
            moredelim=**[is][{\btHL[fill=celadon!40]}]{!}{!}
        }
        \lstset{
            language=WAT,
            style=nccode,
            basicstyle=\footnotesize\ttfamily,
            columns=fullflexible,
            breaklines=true
        }
        \begin{lstlisting}
!i32.const <n>!
!i32.add!

        \end{lstlisting}
    \end{minipage}
\end{minipage}
\end{code}

We look at the cases that yield a small number of variants. There is no direct correlation between the number of identified blocks and the number of unique variants. We manually analyze programs that include a significant number of pure blocks, for which \tool generates few variants. 
We identify two main challenges for diversification.

\emph{1) Constant computation}  We have observed that Souper searches for a constant replacement for more than $45\%$ of the blocks of each program while constant values cannot be inferred. For instance,  constant values cannot be inferred for memory load operations because \tool is oblivious to a memory model. 

% candidates overlapping
\emph{2) Combination computation}  The overlap between code replacements, discussed in \autoref{workflow:generation}, is a second factor that limits the number of unique variants. \tool can generate a high number of variants, but not all replacement combinations are necessarily unique.

\blue{Regarding the potential size overhead of the generated variants, we have compared the \wasm binary size of the 239 programs with their variants. The ratio of size change between the original program and the variants ranges from 82\% (variants are smaller) to 125\% (variants are larger) for all Rosetta programs. This limited impact on the binary size of the variants is good news because they are meant to be distributed to browsers over the network.}

\begin{tcolorbox}[title=Answer to RQ1]

\tool is able to generate diverse variants of \wasm programs for $\nProgramsRosettaDiversified{}/\nProgramsRosetta{}\,(\nProgramsRosettaDiversifiedPercent{})$ programs in our corpus. We observe that programs that include bounded loops and arithmetic expressions are highly prone to diversification. V8's TurboFan compilation to x86 code preserves \nPreservedPercent{} of the transformations performed by \tool.
To our knowledge, this is the first ever realization of automated diversification for \wasm. 

\end{tcolorbox}
 
\subsection{\rqdynamic}\label{results:dynamic}
% recall protocol very shortly
Now, we focus on the 41 programs that have at least 9 unique \wasm variants in order to study the diversity of execution traces. 
% How again
We apply the protocol described in \autoref{sec:protocol-rq2} by executing the \wasm programs and their unique variants  in order to collect the stack operation traces. Then, we compare the traces of each pair of original program and a variant. We run $1906$ program executions and we perform $98774$ trace pair comparisons.

\begin{table*}
\setlength\minrowclearance{1.0pt}
\renewcommand{\arraystretch}{1.4}
\centering
\noindent\begin{minipage}{0.48\linewidth}
\resizebox{\linewidth}{!}{
\begin{tabular}[h]{ l l r r  r r r r r }
 \multicolumn{4}{c}{} &  \multicolumn{5}{c}{} \\
    \midrule
    & NAME & \#var & $\Sigma$ & Min  & Max & Median  & 0 \% & $>$ 0 \%  \\
    \hline
   1 & \textbf{p96} & 220 & 15  &0 & 24062 & 820 &   0.30 & \cellcolor{celadon!60} 99.70 \\
\hline
2 & p56 & 192 & 36  &0 & 45420 & 1416 &   1.84 & \cellcolor{celadon!60} 98.16 \\
\hline
3 & p78 & 159 & 35  &0 & 20501 & 759 &   1.52 & \cellcolor{celadon!60} 98.48 \\
\hline
4 & p111 & 144 & 45  &0 & 2114 & 520 &   3.74 & \cellcolor{celadon!60} 96.26 \\
\hline
5 & \textbf{p166} & 101 & 152  &0 & 44538 & 66 &   45.80 & \cellcolor{celadon!60} 54.20 \\
\hline
6 & p122 & 91 & 34  &0 & 46026 & 6434 &   0.24 & \cellcolor{celadon!60} 99.76  \\
\hline
7 & p67 & 89 & 77  &0 & 94036 & 85692 &   0.29 & \cellcolor{celadon!60} 99.71  \\
\hline
8 & p68 & 85 & 10  &0 & 10554 & 260 &   3.64 & \cellcolor{celadon!60} 96.36 \\
\hline
9 & p80 & 78 & 9  &0 & 17238 & 618 &   3.92 & \cellcolor{celadon!60} 96.08 \\
\hline
10 & p204 & 77 & 42  &0 & 36428 & 3356 &   0.33 & \cellcolor{celadon!60} 99.67  \\
\hline
11 & p183 & 76 & 9  &0 & 90628 & 84402 &   0.57 & \cellcolor{celadon!60} 99.43  \\
\hline
12 & p136 & 62 & 70  &0 & 62953 & 58028 &   0.60 & \cellcolor{celadon!60} 99.40  \\
\hline
13 & p167 & 46 & 232  &8 & 888 & 724 &   0.00 & \cellcolor{celadon!60} 100.00  \\
\hline
14 & p226 & 42 & 13  &0 & 90736 & 74476 &   8.26 & \cellcolor{celadon!60} 91.74 \\
\hline
15 & p99 & 38 & 74  &16 & 9936 & 5037 &   0.00 & \cellcolor{celadon!60} 100.00  \\
\hline
16 & p18 & 36 & 7  &0 & 15620 & 145 &   1.10 & \cellcolor{celadon!60} 98.90  \\
\hline
17 & \textbf{p140} & 29 & 17  &0 & 13280 & 172 &   6.59 & \cellcolor{celadon!60} 93.41  \\
\hline
18 & p59 & 27 & 6  &0 & 85390 & 40 &   1.43 & \cellcolor{celadon!60} 98.57  \\
\hline
19 & p199 & 21 & 87  &0 & 27482 & 728 &   4.68 & \cellcolor{celadon!60} 95.32 \\
\hline
20 & \textbf{p91} & 21 & 21  &0 & 50002 & 228 &   43.81 & \cellcolor{celadon!60} 56.19  \\
\hline
21 & p223 & 21 & 115  &16 & 40911 & 632 &   0.00 & \cellcolor{celadon!60} 100.00  \\
\end{tabular}}
\end{minipage}\hfill%
\begin{minipage}{0.48\linewidth}
\resizebox{\linewidth}{!}{
            \begin{tabular}[h]{ l l r r  r r r r r }
 \multicolumn{4}{c}{} &  \multicolumn{5}{c}{} \\
    \midrule
    & NAME & \#var & $\Sigma$ & Min  & Max & Median  & 0 \% & $>$ 0 \%  \\
    \hline
22 & p168 & 20 & 6  &0 & 22200 & 18896 &   2.20 & \cellcolor{celadon!60} 97.80  \\
\hline
23 & p174 & 18 & 40  &6 & 6566 & 6395 &   0.00 & \cellcolor{celadon!60} 100.00 \\
\hline
24 & \textbf{p81} & 17 & 86  &0 & 4419 & 0 &   84.62 &  15.38 \\
\hline
25 & p141 & 17 & 6  &8 & 2894 & 132 &   0.00 & \cellcolor{celadon!60} 100.00  \\
\hline
26 & p108 & 16 & 6  &0 & 85168 & 79903 &   8.97 & \cellcolor{celadon!60} 91.03 \\
\hline
27 & p98 & 15 & 4  &0 & 33 & 25 &   6.06 & \cellcolor{celadon!60} 93.94  \\
\hline
28 & p89 & 14 & 45  &10 & 15952 & 89 &   0.00 & \cellcolor{celadon!60} 100.00  \\
\hline
29 & p36 & 14 & 52  &312 & 33266 & 30298 &   0.00 & \cellcolor{celadon!60} 100.00 \\
\hline
30 & \textbf{p135} & 13 & 5  &0 & 20288 & 20163 &   3.57 & \cellcolor{celadon!60} 96.43 \\
\hline
31 & p161 & 12 & 91  &240 & 9792 & 1056 &   0.00 & \cellcolor{celadon!60} 100.00 \\
\hline
32 & p147 & 12 & 32  &0 & 54071 & 21274 &   7.14 & \cellcolor{celadon!60} 92.86 \\
\hline
33 & p11 & 10 & 38  &29798 & 51846 & 35119 &   0.00 & \cellcolor{celadon!60} 100.00  \\
\hline
34 & p125 & 10 & 51  &0 & 4399 & 4368 &   7.14 & \cellcolor{celadon!60} 92.86 \\
\hline
35 & p131 & 9 & 4  &140 & 1454 & 685 &   0.00 & \cellcolor{celadon!60} 100.00 \\
\hline
36 & p69 & 9 & 48  &28 & 29243 & 28956 &   0.00 & \cellcolor{celadon!60} 100.00 \\
\hline
37 & p134 & 9 & 20  &4 & 514 & 186 &   0.00 & \cellcolor{celadon!60} 100.00 \\
\hline
38 & \textbf{p74} & 9 & 19  &126 & 8332 & 6727 &   0.00 & \cellcolor{celadon!60} 100.00  \\
\hline
39 & p79 & 9 & 97  &4 & 29 & 16 &   0.00 & \cellcolor{celadon!60} 100.00 \\
\hline
40 & p33 & 9 & 52  &4 & 2342 & 15 &   0.00 & \cellcolor{celadon!60} 100.00  \\
\hline
41 & p157 & 9 & 64  &36 & 242 & 166 &   0.00 & \cellcolor{celadon!60} 100.00 \\
\hline
\\
\end{tabular}}
\end{minipage}
\caption{Dynamic diversity for 41 diversified WASM programs. The dynamic diversity is captured by \DTW{} between traces. The rows are sorted by the number of unique variants per program. The table is structured as follows: the first, second and third columns contain the program id, the number of unique variants and the overall sum of all blocks replacements respectively. Following, the stats for the \DTW{} metric. The colorized cells in the $> 0\%$ column represent high diversification.}\label{tables:rq2:dynamic}
\end{table*}

%\todo{add row "Total" to have the 4 overall frequency of identical and different traces for stack and memory, "-" for the others columns (min/mean/max)}

% explain the figure
Table \ref{tables:rq2:dynamic} summarizes the observed trace diversity, as captured by \DTW{}  (\autoref{metric:stack}), among each program and their variants. The table is structured as follows: the first, second and third columns contain the program id, the number of unique variants and the overall sum of all blocks replacements respectively. The table summarizes the distribution of distances between stack operation trace pairs: the minimum value, the maximum value, the median value, the percentage of values equal to zero and the percentage of values greater than zero. The programs are sorted with respect to the number of unique variants. The green highlight color in $> 0\%$ columns represents more than 50\% of non-zero comparisons, \ie high diversification.  
For instance, the first row shows the trace diversity for \texttt{p96}, where 99.70\% of the pairwise comparisons between all collected traces have a different \DTW{}.

% General stats 
For the stack operation traces, all programs have at least one variant that produces a trace different from the original. All but one (\texttt{p81}) programs have the majority of variants producing a different stack operation trace. This shows the real effectiveness of \tool for diversifying stack operation traces.

% relation between constant and stack operation traces
We manually analyze variants with high and low trace diversity. We observe that constant inferring is effective at changing the stack operation trace. For instance, for program \texttt{p74} shown in \autoref{babbage_code},  \tool removes a loop by replacing it with a constant assignment. The execution of this variant produces traces that are different because the loop pattern is not visible anymore in the trace, and consequently, the distance between the original and the variant traces is large. 

% relation main factor for trace diversity: constant inference
We note that there is no relation between the trace distance and the number of block replacements. A high trace distance does not necessarily imply a high number of  replacements. For instance, program \texttt{p135} has only 4 possible replacements overall its 5 identified blocks yet a median \DTW{} of 20163.

\begin{code}[t]
\centering
\captionof{lstlisting}{Statically different \wasm replacements with the same behavior, gray for the original code, green for the replacement.}\label{rq2:example:lt}
\noindent\rule{\linewidth}{0.4pt}
\noindent\begin{minipage}{.23\linewidth}
\lstdefinestyle{nccode2}{
    numbers=none,
    firstnumber=1,
    stepnumber=1,
    numbersep=10pt,
    tabsize=4,
    showspaces=false,
    breaklines=true, 
    showstringspaces=false,
    moredelim=**[is][{\btHL[fill=black!10]}]{`}{`},
    moredelim=**[is][{\btHL[fill=celadon!40]}]{!}{!}
}
    \lstset{
        language=WAT,
        style=nccode2,
        basicstyle=\footnotesize\ttfamily,
        columns=fullflexible,
        breaklines=true
    }
    \begin{lstlisting}
(1) `i32.lt_u`
(2) `i32.le_s`
    \end{lstlisting}
\end{minipage}\hfill%
\noindent\begin{minipage}{0.2\linewidth}
\lstdefinestyle{nccode2}{
    numbers=none,
    firstnumber=1,
    stepnumber=1,
    numbersep=10pt,
    tabsize=4,
    showspaces=false,
    breaklines=true, 
    showstringspaces=false,
    moredelim=**[is][{\btHL[fill=black!10]}]{`}{`},
    moredelim=**[is][{\btHL[fill=celadon!40]}]{!}{!}
}

    \lstset{
        language=WAT,
        style=nccode2,
        basicstyle=\footnotesize\ttfamily,
        columns=fullflexible,
        breaklines=true
    }
    \begin{lstlisting}
!i32.lt_s!
!i32.lt_u!
    \end{lstlisting}
\end{minipage}\hfill%
\noindent\begin{minipage}{.3\linewidth}
\lstdefinestyle{nccode2}{
    numbers=none,
    firstnumber=1,
    stepnumber=1,
    numbersep=10pt,
    tabsize=4,
    showspaces=false,
    breaklines=true, 
    showstringspaces=false,
    moredelim=**[is][{\btHL[fill=black!10]}]{`}{`},
    moredelim=**[is][{\btHL[fill=celadon!40]}]{!}{!}
}

    \lstset{
        language=WAT,
        style=nccode2,
        basicstyle=\footnotesize\ttfamily,
        columns=fullflexible,
        breaklines=true
    }
    \begin{lstlisting}
(3) `i32.ne`
(4) `local.get 6`
    \end{lstlisting}
\end{minipage}\hfill%
\noindent\begin{minipage}{0.2\linewidth}
\lstdefinestyle{nccode2}{
    numbers=none,
    firstnumber=1,
    stepnumber=1,
    numbersep=10pt,
    tabsize=4,
    showspaces=false,
    breaklines=true, 
    showstringspaces=false,
    moredelim=**[is][{\btHL[fill=black!10]}]{`}{`},
    moredelim=**[is][{\btHL[fill=celadon!40]}]{!}{!}
}

    \lstset{
        language=WAT,
        style=nccode2,
        basicstyle=\footnotesize\ttfamily,
        columns=fullflexible,
        breaklines=true
    }
    \begin{lstlisting}
!i32.lt_u!
!local.get 4!
    \end{lstlisting}
\end{minipage}
\end{code}

% low stack diversity cases

We subsequently analyze the cases where diversification is not reflected in stack operation traces. For example, more than 40\% of the pairwise \DTW{} distances for \texttt{p166}, \texttt{p91} and \texttt{p81} are equal to zero. This indicates a lower diversity among the population of variants, than for all the other programs. This happens because some variants have two different bitcode instructions (original and replacement) that trigger the same stack operations. The instructions in \autoref{rq2:example:lt} are concrete cases of such kind of replacements. 
The four cases in \autoref{rq2:example:lt} leave the same value in the stack operation trace. For each case, the original instruction and the replacement are semantically equal in the program domain. The fourth case is a local variable index reallocation, this replacement only changes the index of the local variable but not the event in the stack operation trace.
These replacements are sound, produce statically diverse code, but they are not useful to dynamically diversify the original program. This confirms the complementary of using static and dynamic metrics to assess diversification.

% significance for security
The effectiveness of \tool on diversifying stack operation traces is significant. In a security context, such diverse stack operation traces are likely to mitigate  potential side-channel attacks \cite{usenixWASM2020}. Notably, the attacks based on code profiling are affected when the executed opcodes and the corresponding profiles are different~\cite{Rudd2017AddressOC}.

%\todo{TBD, add the size relation between sections for all programs?}

%\blue{The size of the code in \wasm programs directly affects the execution time. We observe the following trends. If CROW can infer control flows as constants in the original program, then the variants execute faster than the original, sometimes by an order of magnitude. On the other hand, if the original block is replaced by a block with more instructions, then the variants tend to run slower than the original. In the web context, the execution time of the code and the size of the served resources (in this case the \wasm binaries) need to be taken into account due to performance, \ie the programs should be fast and small. \tool can be configured to not replace blocks larger than the original avoiding these constraints.}

\begin{tcolorbox}[title=Answer to RQ2]

\tool is successful at generating diverse \wasm variant programs, for which we are able to observe different stack operation traces. In other words, \tool generates dynamically different binaries, and ensures that variants of a given program yield different stack operation traces.
\end{tcolorbox}

\subsection{\rqlibsodium}\label{results:rqlibsodium}
% General stats

We run \tool on each of the $102$ modules of libsodium with a 6-hour timeout. We find
$45/102$ modules that do not contain any pure block, so they are not amenable to our diversification technique.
\tool produces at least one valid \wasm module variant for $15$ of the remaining $57$ modules.

\begin{table*}
    \centering
    \fontsize{8.15pt}{8.15pt}\selectfont
    \lstset{% Break on underscores with lstinline
      literate={\_}{}{0\discretionary{\_}{}{\_}},%
    }
    \renewcommand{\arraystretch}{0.5}
    
    % https://tex.stackexchange.com/a/11215
    \setlength{\aboverulesep}{0pt}
    \setlength{\belowrulesep}{0pt}
    \begin{tabular}{p{0.385\linewidth}t |l p{0.35\linewidth}r}
        \toprule
        \textbf{Module \& Description} & \textbf{\#var} & \textbf{\#func} & \textbf{Diversified Functions}  & \textbf{\#calls}\\
        \midrule
        \textbf{\texttt{argon2-core}}\newline Core functions for the implementation of the Argon2 key derivation (hash) function~\cite{biryukov2016argon2}.
& $17$ & $6$
& \begin{tabular}[t]{@{}l}
{\texttt{argon2\_finalize}}\\{\texttt{argon2\_free\_instance}}\\{\texttt{argon2\_initialize}}
\end{tabular}
& \begin{tabular}[t]{r@{}}
{$0$}\\{$0$}\\{$0$}
\end{tabular}\\
\hline
\textbf{\texttt{argon2-encoding}}\newline Functions for encoding and decoding (including salting) Argon2~\cite{biryukov2016argon2} hash strings.
& $11$ & $2$
& \begin{tabular}[t]{@{}l}
{\texttt{argon2\_decode\_string}}\\{\texttt{argon2\_encode\_string}}
\end{tabular}
& \begin{tabular}[t]{r@{}}
{$0$}\\{$0$}
\end{tabular}\\
\hline
\textbf{\texttt{blake2b-ref}}\newline Reference implementation for the BLA\-KE2~\cite{aumasson2013blake2} hash function.
& $7$ & $11$
& \begin{tabular}[t]{@{}l}
{\texttt{blake2b}}\\{\texttt{blake2b\_salt\_personal}}\\{\texttt{blake2b\_update}}
\end{tabular}
& \begin{tabular}[t]{r@{}}
{$0$}\\$1.46\mathrm{E}{+04}$\\$2.04\mathrm{E}{+04}$
\end{tabular}\\
\hline
\textbf{\texttt{chacha20\_\allowbreak{}ref}}\newline Reference implementation of the ChaCha20 stream cipher~\cite{bernstein2008chacha}.
& $7$ & $5$
& \begin{tabular}[t]{@{}l}
{\texttt{chacha20\_encrypt\_bytes}}\\{\texttt{stream\_ietf\_ext\_ref\_xor\_ic}}\\{\texttt{stream\_ref}}\\{\texttt{stream\_ref\_xor\_ic}}
\end{tabular}
& \begin{tabular}[t]{r@{}}
$3.51\mathrm{E}{+06}$\\$7.62\mathrm{E}{+03}$\\$1.14\mathrm{E}{+04}$\\$1.14\mathrm{E}{+05}$
\end{tabular}\\
\hline
\textbf{\texttt{codecs}}\newline Implementations of commonly used codecs for conversions between binary formats like Base64~\cite{josefsson2006base16}.
& $79$ & $5$
& \begin{tabular}[t]{@{}l}
{\texttt{sodium\_base642bin}}\\{\texttt{sodium\_base64\_encoded\_len}}\\{\texttt{sodium\_bin2base64}}\\{\texttt{sodium\_bin2hex}}\\{\texttt{sodium\_hex2bin}}
\end{tabular}
& \begin{tabular}[t]{r@{}}
{$0$}\\{$0$}\\{$0$}\\$2.57\mathrm{E}{+05}$\\{$0$}
\end{tabular}\\
\hline
\textbf{\texttt{core\_\allowbreak{}ed25519}}\newline Implementation of the Edwards-curve Digital Signature Algorithm~\cite{bernstein2012high}.
& $2$ & $19$
& \begin{tabular}[t]{@{}l}
{\texttt{crypto\_core\_ed25519\_is\_valid\_point}}
\end{tabular}
& \begin{tabular}[t]{r@{}}
{$0$}
\end{tabular}\\
\hline
\textbf{\texttt{crypto\_\allowbreak{}scrypt-common}}\newline Utility and low-level API functions for the scrypt key derivation (hash) function~\cite{percival2009stronger}.
& $5$ & $5$
& \begin{tabular}[t]{@{}l}
{\texttt{escrypt\_gensalt\_r}}
\end{tabular}
& \begin{tabular}[t]{r@{}}
{$0$}
\end{tabular}\\
\hline
\textbf{\texttt{pbkdf2-sha256}}\newline Implementation of the Password-Based Key Derivation Function 2 (PBKDF2)~\cite{RFC2898}.
& $14$ & $1$
& \begin{tabular}[t]{@{}l}
{\texttt{escrypt\_PBKDF2\_SHA256}}
\end{tabular}
& \begin{tabular}[t]{r@{}}
{$0$}
\end{tabular}\\
\hline
\textbf{\texttt{pwhash\_\allowbreak{}scryptsalsa208sha256}}\newline High-level API for the scrypt key derivation function~\cite{percival2009stronger}.
& $8$ & $19$
& \begin{tabular}[t]{@{}l}
{\texttt{crypto\_pwhash\_scryptsalsa208sha256}}
\end{tabular}
& \begin{tabular}[t]{r@{}}
{$0$}
\end{tabular}\\
\hline
\textbf{\texttt{pwhash\_\allowbreak{}scryptsalsa208sha256\_\allowbreak{}nosse}}\newline Same as above, but does not use Streaming SIMD Extensions (SSE).
& $32$ & $3$
& \begin{tabular}[t]{@{}l}
{\texttt{escrypt\_kdf\_nosse}}\\{\texttt{salsa20\_8}}
\end{tabular}
& \begin{tabular}[t]{r@{}}
{$0$}\\{$0$}
\end{tabular}\\
\hline
\textbf{\texttt{randombytes}}\newline Pseudorandom number generators.
& $1$ & $11$
& \begin{tabular}[t]{@{}l}
{\texttt{randombytes\_uniform}}
\end{tabular}
& \begin{tabular}[t]{r@{}}
$5.61\mathrm{E}{+02}$
\end{tabular}\\
\hline
\textbf{\texttt{salsa20\_\allowbreak{}ref}}\newline Contains a reference implementation of the Salsa20 stream cipher~\cite{bernstein2008salsa20}.
& $12$ & $2$
& \begin{tabular}[t]{@{}l}
{\texttt{stream\_ref}}\\{\texttt{stream\_ref\_xor\_ic}}
\end{tabular}
& \begin{tabular}[t]{r@{}}
$1.14\mathrm{E}{+04}$\\$1.14\mathrm{E}{+05}$
\end{tabular}\\
\hline
\textbf{\texttt{scalarmult\_\allowbreak{}ristretto255\_\allowbreak{}ref10}}\newline Implementation of the Ristretto255 prime order elliptic curve group~\cite{ristretto}.
& $29$ & $4$
& \begin{tabular}[t]{@{}l}
{\texttt{scalarmult\_ristretto255}}\\{\texttt{scalarmult\_ristretto255\_base}}\\{\texttt{scalarmult\_ristretto255\_scalarbytes}}
\end{tabular}
& \begin{tabular}[t]{r@{}}
{$0$}\\{$0$}\\{$0$}
\end{tabular}\\
\hline
\textbf{\texttt{stream\_\allowbreak{}chacha20}}\newline High-level API for the ChaCha20 stream cipher~\cite{bernstein2012high}.
& $2$ & $15$
& \begin{tabular}[t]{@{}l}
{\texttt{crypto\_stream\_chacha20}}\\{\texttt{crypto\_stream\_chacha20\_ietf}}\\{\texttt{crypto\_stream\_chacha20\_ietf\_ext}}\\{\texttt{crypto\_stream\_chacha20\_ietf\_ext\_xor\_ic}}\\{\texttt{crypto\_stream\_chacha20\_ietf\_xor}}\\{\texttt{crypto\_stream\_chacha20\_ietf\_xor\_ic}}\\{\texttt{crypto\_stream\_chacha20\_xor}}\\{\texttt{crypto\_stream\_chacha20\_xor\_ic}}
\end{tabular}
& \begin{tabular}[t]{r@{}}
$6.65\mathrm{E}{+02}$\\$3.19\mathrm{E}{+03}$\\$2.66\mathrm{E}{+03}$\\$1.68\mathrm{E}{+02}$\\$1.68\mathrm{E}{+02}$\\$2.32\mathrm{E}{+03}$\\{$0$}\\$1.68\mathrm{E}{+02}$
\end{tabular}\\
\hline
\textbf{\texttt{verify}}\newline Functions used to compare secrets in constant time to avoid timing attacks.
& $7$ & $6$
& \begin{tabular}[t]{@{}l}
{\texttt{crypto\_verify\_16}}\\{\texttt{crypto\_verify\_32}}\\{\texttt{crypto\_verify\_64}}
\end{tabular}
& \begin{tabular}[t]{r@{}}
$2.69\mathrm{E}{+05}$\\$3.40\mathrm{E}{+03}$\\{$0$}
\end{tabular}\\
        & & & \\ % Produces a tiny empty line, maybe there is a better way to do this ;D
        \textbf{Total} & $\mathbf{256}$ & $\mathbf{114}$ & \textbf{$\mathbf{40}$ functions} \\
        \bottomrule
    \end{tabular}%
    
    \caption[Diversified libsodium modules]{%
    Libsodium modules with at least one variant generated by \tool.
    The columns on the left include the facts about each module.
    The first column contains the name and the functional description of the modules.
    The second column, \textit{\#var} (highlighted) gives the number of unique variants generated by \tool.
    The third column, \textit{\#func}, lists the total amount of  functions in each module.
    The remaining columns include a list of functions that \tool has successfully diversified and the number of calls per function in the test suite.}\label{tables:libsodium}
\end{table*}

% Table description
Table \ref{tables:libsodium} presents the key results for these $15$ successfully diversified modules.
The first two columns contain the name and description of the diversified module, and, the number of unique static variants.
The other columns show the total number of functions inside the module, the names of the diversified functions and the number of calls to each function in the considered tests.
%For instance, the third row contains the description for the \texttt{blake2b-ref} module for which we produce $8$ unique variants. 
%That module contains $11$ functions and we generate variants for $3$ functions, listed in the right column.
%Out of those $3$ functions, \texttt{blake2b} is not called by the tests, and the other two are called more than $10$k times. 
%Overall, $19$ functions are executed by the considered tests.

% How many we can produce
\emph{Generation of \wasm  library variants from \wasm module variants.}
The successfully diversified modules can be combined to obtain a large pool of different versions of the packaged libsodium \wasm library. The Cartesian product of all module variants produces in theory $1.66\mathrm{E}{+15}$ unique libsodium variants.
% How do we pick variants for evaluation?
Yet, it is unpractical to store and execute this large number of variants.
Thus, we sample the pool of possible variants to evaluate our generated variants.
First, for each of the $256$ modules, we rank each module variant with respect to the number of lines changed in the final \wasm textual format.
Then, to produce the $i$-th library variant, we combine the $i$-th variant for each module of libsodium, in order to produce maximally diversified library variants first.
If a module has less than $i$ variants, we use the original, non-diversified module.
According to \autoref{tables:libsodium}, the maximum number of unique variants for a single module is $79$ (\texttt{codecs} module). Thus, we sample $79$ unique libsodium variants, ordered by the amount of diversification (the first variant contains the most changes, and so on).
For each variant we execute the complete test suite to validate its correctness.
All test cases successfully pass for all diversified library binaries.

\begin{figure}
    \centering
      \includegraphics[width=\linewidth]{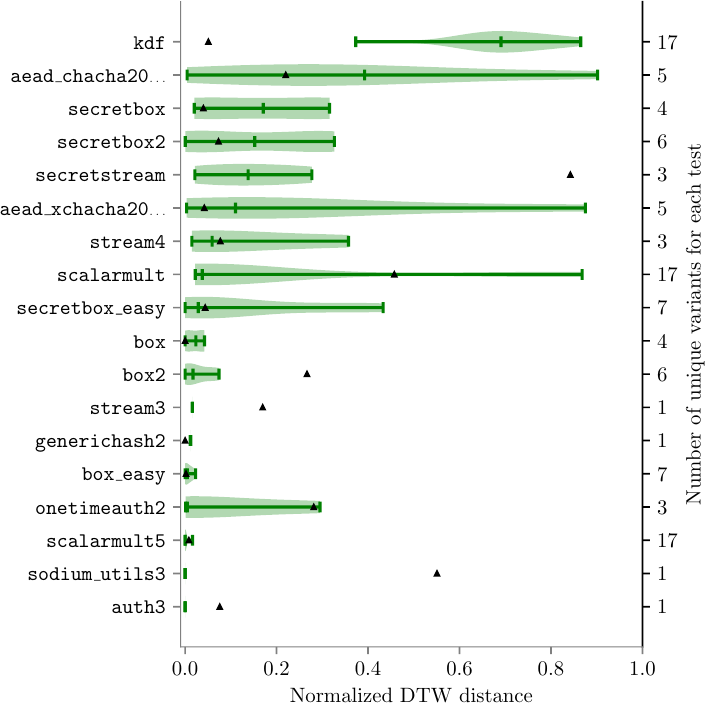}
      \caption{%
      Distribution of normalized \DTW{} distances over the set of libsodium variants covered by each test.
      The left Y axis lists the name of each test.
      The number of unique variants used per test is listed on the right Y axis.
      The black triangles point to the \DTW{} distance between two different stack operation traces of the original test with different random seeds.
      }
      \label{fig:rqlibsodium}
\end{figure}

\textbf{Dynamic evaluation of libsodium variants.}
% Plots: DTW distances of stack operation traces
We compare the dynamic behaviour of the original libsodium and the $79$ library variants.
Figure \ref{fig:rqlibsodium} illustrates the distribution of \DTW{} of all collected traces for each libsodium test.
The \DTW{} distance is calculated between each diversified trace and the corresponding original trace for the same test.
Each horizontal bar gives the distribution of \DTW{} over the $79$ diversified libraries per test.
The black triangles show the \DTW{} distance between two different executions of the same test with different random seeds. They serve as a baseline to compare the artificial diversity introduced by \tool, against the natural trace diversity that appears because of random number generation.

% Overall findings
For $18/19$ tests, we observe that \tool's diversified modules produce a different trace than the original.
The wider violin plots that reach the right-hand side of the figure include variants that significantly diversify the test execution.
We observe that $4/18$ tests stand out as they include variants with at least $0.8$ normalized \DTW{} distance.
For $6/18$ tests, there is a medium trace diversity as their \DTW{} distributions lie in the mid/left side of the plot.
For the rest $8/18$ tests we observe a significantly smaller \DTW{} distance.

This means that, in the context of this cryptographic library, \tool is able to find variants that have a huge impact on  the dynamic stack behaviour of the program.
Meanwhile, some other replacements can have only a marginal impact during the operation of the program.
One factor that can affect this is the ``centrality'' of the code that is being replaced.
Diversified code that is called often, potentially inside loops, will have a greater impact on the stack trace of a program compared to code that is only called, for example, only during the initialization of the program.

% Compared to "natural diversity"
When we compare the trace diversity against the diversity due to pseudo-number generation (black triangles in \autoref{fig:rqlibsodium}), we observe that:
for $2/18$ tests \tool trace diversification is always larger than the one due to random number generation,
for $11/18$ tests there exist some variants that exhibit larger trace diversification than random number generation
and for $5/18$ tests \tool trace diversification is always smaller than the one due to random number generation.

% answer to RQ libsodium
\begin{tcolorbox}[title=Answer to RQ3]

We have successfully applied  \tool to libsodium, one of the leading \wasm cryptography libraries.
We have shown that \tool is able to create statically different variants of this real-world library, all of which being distributable to users.
Our original experiments to measure the trace diversity of libsodium have proven that the generated variants exhibit significantly different execution traces compared to the original non-diversified libsodium binary.
The take-away of this experiment is that \tool works on complex code. 

\end{tcolorbox}

%%%%%%%%%%%%%%%%%%%%%%%%%%%%%%%%%%%%%%%%%%%%%%%%%%
\section{Threats to Validity}\label{sec:discussion}

\emph{Internal:}
The timeout in the exploration stage is a determinant factor to generate unique variants. It is required to bound the experimental time. If the timeout is increased, the number of variants and unique variants might increase. 

%\tool and the evaluation scripts may contain bugs, since we have modified and extended four large and complex tools (LLVM, Souper, SWAM, STRAC) to achieve our goal.
%To encourage the replication of our work and for the sake of open science, our data and code is freely available at \url{https://github.com/KTH/slumps}.

\emph{External:}
The \nProgramsRosetta{} programs in our Rosetta corpus may not reflect the constructs used in the \wasm programs in the wild. Yet our experiment on libsodium shows that the results on the Rosetta corpus hold on real code. To increase external validity, we hope to see more benchmarks of \wasm programs published by the research community.

\emph{Scale:}
% Study subjects and dynamic comparison:
We measure behavioral diversity with DTW. We are aware that this behavioral diversity metric does not scale infinitely. To make comparisons between  large execution traces, it may be necessary to use a more scalable metric. To mitigate this scale problem in future work, one option is to compare software traces using entropy analysis, as proposed by Miranskyy \etal \cite{Miranskyy2012}. 

\section{Related Work}\label{sec:related-work}
Program diversification approaches can be applied at different stages of the development pipeline.

% Binary code diversification
\emph{Static diversification:}
This kind of diversification consists in synthesizing, building and distributing different, functionally equivalent, binaries to end users. This aims at increasing the complexity and applicability of an attack against a large population of users~\cite{cohen1993operating}. Jackson \etal~\cite{jackson2011compiler} argue that the compiler can be placed at the heart of the solution for software diversification;
they propose the use of multiple semantic-preserving transformations to implement massive-scale software diversity in which each user gets their own diversified variant.
Dealing with code-reuse attacks, Homescu \etal~\cite{homescu2013profile} propose inserting NOP instruction directly in LLVM IR to generate a variant with different code layout at each compilation. 
In this area, Coppens \etal~\cite{coppens2013feedback} use compiler transformations to iteratively diversify software.
The aim of their work is to prevent reverse engineering of security patches for attackers targeting vulnerable programs.
Their approach, continuously applies a random selection of predefined transformations using a binary diffing tool as feedback.
A downside of their method is that attackers are, in theory, able to identify the type of transformations applied and find a way to ignore or reverse them.
Our work can be extended to address this issue, providing a synthesizing solution which is more general than specific transformations.

% The Superdiversifier
The work closest to ours is that by Jacob \etal~\cite{jacob2008superdiversifier}.
These authors propose the use of a ``superdiversification'' technique, inspired by superoptimization~\cite{1987_Massalin_Sueroptimizer},
to synthesize individualized versions of programs.
In the work of Massalin, a superoptimizer aims to synthesize the shortest instruction sequence that is equivalent to the original given sequence.
On the contrary, the tool developed by Jacob \etal does not output only the shortest instruction sequence, but any sequences that implement the input function.
This work focuses on a specific subset of X86 instructions.
Meanwhile, our approach works directly with LLVM IR, enabling it to generalize to more languages and CPU architectures.
Specifically, we apply our tool on \wasm, something not possible with the X86-specific approach of that paper.

\emph{Runtime diversification:}
Previous works have attempted to generate diversified variants that are alternated during execution.
It has been shown to drastically increase the number of execution traces that a side-channel attack requires to succeed.
Amarilli \etal~\cite{amarilli2011can} are the first to propose generation of code variants against side-channel attacks.
Agosta \etal~\cite{agosta2015meet} and Crane \etal~\cite{crane2015thwarting}
modify the LLVM toolchain to compile multiple functionally equivalent variants to randomize the control flow of software,
while Courouss{\'e} \etal~\cite{courousse2016runtime} implement an assembly-like DSL to generate equivalent code at runtime in order to increase protection against side-channel attacks.
\tool focuses on static diversification of software. However, because of the specificities of code execution in the browser, this is not far from being a dynamic approach. Since \wasm is served at each page refreshment, every time a user asks for a \wasm binary, she can be served a different variant provided by \tool.

\section{Conclusion}
\label{sec:conclusion}

Security has been a major driver for the design of \wasm. Diversification is one additional protection mechanism that has been not yet realized for it. In this paper, we have presented \tool, the first code diversification approach for \wasm.
We have shown that \tool is able to generate variants for a large variety of programs, including a real-world cryptographic library. Our original experiments have comprehensively assessed the generated diversity:  we have shown that \tool generates diversity both among the binary code variants as well as in the execution traces collected when executing the variants.
Also, we have successfully observed diverse execution traces for the considered cryptographic library, which can protect it against a range of side channel attacks.

Future work includes increasing the number of unique variants that are generated, by working on block replacement overlapping detection.  Also, the exploration stage and the identification of code replacements is a highly parallelizable process,  this would increase diversification performance in order to meet the demands of the internet scale.

\section*{Acknowledgement}
 This work has been partially supported by the WASP program, and by the TrustFull project financed by the Swedish Foundation for Strategic Research. We would like to thank John Regehr, the Souper's team, and the Fastly team for their support.

\bibliographystyle{IEEEtranS}
% we use one single BIB file to be able to quickly find existing entries with a single Ctrl-F command
\IEEEtriggeratref{22} % Recommended way to balance columns is NOT with the balance package because it splits references in half
\bibliography{main}

\end{document}